\documentclass[letterpaper, 10 pt, conference]{ieeeconf}  %

\IEEEoverridecommandlockouts                              %
\usepackage{graphics} %
\usepackage{graphicx}
\usepackage{xcolor}
\usepackage{epsfig} %
\usepackage{times} %

\usepackage{amssymb,amsmath,amsthm}
\usepackage{mathtools}
\usepackage{dsfont}
\usepackage{balance}

\usepackage{multicol}
\usepackage{multirow}
\usepackage{booktabs}
\usepackage{rotating}
\usepackage{tabularx}
\usepackage{adjustbox}
\usepackage{hyperref}
\usepackage{url}
\usepackage{tikz} %
\usepackage[normalem]{ulem}

\usepackage{xfrac}

\theoremstyle{definition}
\newtheorem{proposition}{Proposition}

\newtheorem{remark}{Remark}

\title{\LARGE \bf
Time-To-Reach Separation and Safety Filtering for \\
Safe, Fair, and Efficient Multi-Agent Coordination
}

\newif\ifarxiv
\arxivtrue %

\author{Matthew Low$^1$, Jasmine Jerry Aloor$^2$, Victoria Marie Tuck$^3$, Pierluigi Nuzzo$^1$ and Jason J. Choi$^4$%
\thanks{This work was supported in part by the National Science Foundation under awards 2514748 and 2448886.}%
\thanks{$^1$M. Low and P. Nuzzo are with the Department of Electrical Engineering and Computer Sciences, University of California, Berkeley. $^2$J.J. Aloor is with the Department of Aeronautics and Astronautics, Massachusetts Institute of Technology. $^3$V. Tuck is with the GRASP Laboratory, University of Pennsylvania. $^4$J.J. Choi is with the Department of Electrical and Computer Engineering, University of California, Los Angeles. {\tt\small m.low@berkeley.edu, jjhchoi@ucla.edu.} AI tools (OpenAI ChatGPT, Anthropic Claude, Google Gemini) have been used to improve the clarity of the manuscript.}%
\thanks{Full version with appendix: \href{https://arxiv.org/abs/arXiv:2605.20625}{\texttt{arXiv:2605.20625}}. Source code: \url{https://github.com/mlow2/ttr_guidance}}%
}

\begin{document}

\newcommand{\cbf}{h}
\newcommand{\traj}{\mathrm{x}}
\newcommand{\ctrl}{\mathrm{u}}
\newcommand{\trajeta}{\eta}
\newcommand{\trajpsi}{\psi}
\newcommand{\xquery}{x_{\text{query}}}
\newcommand{\Lx}{L^{x}}
\newcommand{\Lout}{L^{\text{out}}}
\newcommand{\Lin}{L^{\text{in}}}
\newcommand{\Lio}{L^{\text{io}}}
\newcommand{\ball}{\text{B}}
\newcommand{\rect}{\text{Rect}}
\newcommand{\pibackup}{\pi^{\text{backup}}}
\newcommand{\pisafety}{\pi^{\text{safe}}}
\newcommand{\piexploration}{\pi^{\text{exp}}}
\newcommand{\pisafeexploration}{\pi}
\newcommand{\safetythreshold}{\epsilon}
\newcommand{\ntraj}{n^{\textnormal{traj}}}
\newcommand{\niter}{n^{\textnormal{iter}}}

\newcommand{\trajinduced}{\zeta}
\newcommand{\target}{l}
\newcommand{\unsafefcn}{g}
\newcommand{\constraintfcn}{c}

\newcommand{\unsafeset}{\mathcal{X}_{\mathcal{U}}}
\newcommand{\targetset}{\mathcal{T}}
\newcommand{\constraintset}{\mathcal{C}}
\newcommand{\safeset}{\mathcal{S}\hspace{-1pt}a\!f\!e}
\newcommand{\downstream}{\mathcal{D}}
\newcommand{\dist}{\text{dist}}
\newcommand{\ttr}{T_{\min}}
\newcommand{\ttrdelay}{T_{\text{delay}}}
\newcommand{\ttrdelayptwo}[1][i]{T^{#1}_{\text{delay}\scalebox{0.5}{$+$}}}
\newcommand{\ttrtarget}{T_{\text{target}}}
\newcommand{\clip}{\text{clip}}
\newcommand{\cfset}{\mathcal{U}}
\newcommand{\pcfset}{\mathcal{U}_{\ge0}}

\newcommand{\norm}[1]{\left\lVert #1\right\rVert}
\newcommand{\set}[1]{\left\{ #1\right\}}

\newcommand{\dataset}{\mathcal{D}}

\newcommand{\dVdt}{\frac{\partial V}{\partial t}}
\newcommand{\dVhatdt}{\frac{\partial \widehat{V}}{\partial t}}

\newcommand{\HRule}{\noindent\rule{\linewidth}{0.1mm}\newline}

\newcommand{\R}{\mathbb{R}}

\newcommand{\mat}[1]{\vec{#1}}
\newcommand{\defn}[1]{\emph{#1}}
\newcommand{\xhdr}[1]{{\noindent\bfseries #1}.}

\newcommand{\xdim}{n}
\newcommand{\udim}{m}
\newcommand{\uset}{U}
\newcommand{\datasetu}{\mathcal{D}_{u}}
\newcommand{\datasetv}{\mathcal{D}_{v}}
\newcommand{\tildeU}{\datasetu}
\newcommand{\ctrlsetv}{\datasetv}

\newcommand{\datadrivenV}{\widehat{V}}
\newcommand{\ddV}{\datadrivenV}

\newcommand{\tildeV}{\widetilde{V}}
\newcommand{\tildetraj}{\widetilde{\bm{x}}}
\newcommand{\tildectrl}{\widetilde{\bm{u}}}
\newcommand{\brt}{\mathcal{BRT}}
\newcommand{\brat}{\mathcal{BRAT}}

\newcommand{\ctrlv}{\bm{v}}

\newcommand{\ctrlw}{\bm{w}}
\newcommand{\hattraj}{\widehat{\bm{x}}}
\newcommand{\hatctrl}{\widehat{\bm{u}}}
\newcommand{\hatv}{\widehat{v}}
\newcommand{\hatvi}{\widehat{v}^{i}}

\newcommand{\hatV}{\widehat{V}}

\newcommand{\truev}{\widetilde{v}^i}

\newcommand{\uncertainset}{\mathcal{E}}
\newcommand{\ddh}{\widehat{H}}

\newcommand{\airspeed}{\mathrm{v}}

\newcommand{\hruleopen}{\vspace{0.5em}\hrule\vspace{0.5em} \noindent \;}
\newcommand{\hruleclose}{\hrule \vspace{0.75em}}

\newcommand{\todo}[1]{{\color{blue}{TODO: #1}}}
\newcommand{\jcnote}[1]{{\color{purple}{#1}}}
\newcommand{\ml}[1]{{\textcolor[HTML]{FF8C00}{#1}}} %
\newcommand{\mlq}[1]{{\textcolor[HTML]{7030A0}{#1}}} %
\newcommand{\rev}[1]{{\textcolor[HTML]{C71585}{#1}}} %

\newcommand{\Ppasstwo}{\ensuremath{P_{\text{pass}}^{\scalebox{0.70}{$+$}}}} %

\maketitle
\ifarxiv
  \begin{tikzpicture}[remember picture,overlay]
    \node[anchor=south, yshift=0.75cm] at (current page.south) {
      \parbox{\textwidth}{
        \footnotesize
        Accepted to appear in the 65th IEEE Conference on Decision and Control (CDC 2026).
      }
    };
  \end{tikzpicture}
\fi
\thispagestyle{empty}
\pagestyle{empty}

\begin{abstract}
Advanced Air Mobility operations are expected to significantly increase aerial traffic in urban airspace, requiring autonomous traffic management systems to ensure collision-free operations in highly congested environments. In this paper, we propose a multi-agent coordination framework that uses minimum time-to-reach (TTR) as a unifying metric for priority assignment, temporal separation, and safety filtering. We focus on the problem of coordinating multiple aerial vehicles merging into an air corridor while maintaining safe separation between vehicles. Vehicles are assigned arrival-consistent priority based on TTR, and target TTR values are used to enforce temporal spacing, which  induces spatial separation. A priority-consistent safety filtering layer based on Hamilton-Jacobi reachability value functions promotes collision avoidance while minimally modifying the reference guidance. Simulation results in a highly congested corridor merging scenario show that the proposed method improves safety, fairness, and efficiency compared to time-optimal guidance and priority-agnostic safety filtering.
\end{abstract}

\section{Introduction}

Advanced Air Mobility (AAM) is projected to introduce tens of thousands of daily flights in urban areas, pushing beyond the capacity of current air traffic management~\cite{FAA_AAM_Forecasts2025, FAA_ATC_Hiring2025}. 
Dedicated air corridor designs have been proposed as a structured approach to organizing traffic flows in dense airspace~\cite{homola_corridors_2012, US_DOT2025_AAM_National_strategy, SESARJointUndertaking}. However, when multiple vehicles converge to the same corridor, coordinating their arrival while maintaining safety and efficiency becomes a challenging multi-agent control problem~\cite{wang_quasi-dynamic_2023, aloorDecentralizedCoordinationAutonomous2026}.

Existing approaches to multi-vehicle traffic management and control can broadly be categorized into strategic flow management and tactical conflict resolution. At the strategic level, flow management and scheduling methods assign arrival times or sequencing orders prior to flight. These approaches operate at the operational planning level and are effective in managing throughput and scheduling fairness~\cite{chin_traffic_2023}. At the tactical level, reactive deconfliction systems such as TCAS~\cite{TCAS2011introduction} and ACAS-X~\cite{kochenderfer2012_ACAS} provide decentralized pairwise collision avoidance with formal safety guarantees. These systems are designed primarily for imminent pairwise conflict resolution rather than coordinated multi-vehicle sequencing. 

Overall, prior approaches remain limited for the highly congested air-corridor merging problem considered in this work. Strategic scheduling typically does not account for vehicle dynamics and local and transient airspace congestion. Therefore, the assigned schedules may be dynamically infeasible or require large safety-related deviations that reduce efficiency. Moreover, fairness metrics used in scheduling are often not aligned with safety objectives. On the other hand, purely reactive safety mechanisms are inherently myopic: a collision avoidance maneuver for one pair of vehicles alters the geometry for downstream vehicles, potentially triggering new conflicts. This domino effect has been identified as a fundamental failure mode of decentralized deconfliction in high-density airspace~\cite{wang_quasi-dynamic_2023, zhang_tactical_2023, AloorJ-RSS-25}. During corridor merging, traffic naturally compresses in space, and reactive pairwise avoidance alone can lead to simultaneous conflicts. %

In this paper, we propose a framework that uses the \textit{minimum time-to-reach} (TTR)~\cite{mitchell2002application} as a unifying metric for priority assignment, temporal separation, and integration with safety filtering. Specifically, TTR is used (1) to assign arrival-consistent priority in a fair manner; (2) to create temporal separation by assigning target TTR values, which naturally induces spatial separation as a strategic safety mechanism; and (3) to integrate with a safety filtering layer that enhances the final safe coordination. The key idea is that TTR naturally defines a fair and dynamically feasible arrival ordering: vehicles with smaller TTR can arrive at the corridor sooner and are therefore assigned the right-of-way. This priority rule not only embodies the principle of first-come-first-served fairness~\cite{chin2021efficiency} but also naturally arises from the actual vehicle dynamics via reachability analysis. While TTR has been adopted for single-vehicle guidance~\cite{doshi2022hamilton} and as a reward signal in reinforcement learning~\cite{lyu2020iros, AloorJ-RSS-25}, it has not been directly used as a coordination variable for safe multi-agent coordination.

Prior work that integrates both strategic coordination and tactical safety enforcement for multi-agent coordination is most related to our approach. System-level coordination using formal methods has been studied, integrating ground infrastructure and centralized planning~\cite{bharadwaj_decentralized_2021}. Other work employs reinforcement learning-based approaches for strategic coordination combined with safety filters~\cite{AloorJ-RSS-25, ahmad2025hmarl}. Responsibility allocation for safety filtering has also been formulated as a centralized joint optimization problem~\cite{autenrieb2026combinatorial}. These efforts share a similar hierarchical structure that combines strategic coordination and tactical safety mechanisms. Unlike them, our approach uses instead a dynamically consistent metric, namely, TTR, to jointly determine priority, temporal spacing, and safety.%

\vspace{-0.5em}
\section{Background}
\label{sec:background}
\vspace{-0.25em}
Consider the nonlinear system dynamics
\vspace{-0.25em}
\begin{equation}
    \dot{\traj}(t)\!=\!f(\traj(t), \ctrl(t)) \;\; \text{for} \; t\ge 0, \quad \traj(0) = x, \vspace{-0.25em}
\label{eq:dynsys}
\end{equation}
with initial state $x$, state $\traj(t) \in \mathbb{R}^{\xdim}$, control $\ctrl(t) \in \uset \subset \mathbb{R}^{\udim}$, where $\uset$ is the compact control input set. We assume that $f$ is Lipschitz continuous in the state and $\ctrl(\cdot)$ is Lebesgue measurable for the forward completeness of $\traj(\cdot)$~\cite{sastry2013nonlinear}.

\subsection{Safety and Reachability Analysis}
\label{ssec:reachability}

In this work, we consider two reachability problems: the minimal backward reachable tube (BRT) problem, used to compute the value function for deconfliction safety control, and the backward reach-avoid tube (BRAT) problem, used to compute the optimal control for reaching the corridor and the associated TTR value. We denote the \textit{constraint set} as $\constraintset$ and the \textit{target set} as $\targetset$.

The minimal BRT of the target set $\targetset$ is the set of initial states from which the system reaches the target set $\targetset$ over the time horizon $t$, under \textit{any} control signal $\ctrl$:
\vspace{-0.25em}
\begin{equation*}
        \brt(t; \targetset) \! \coloneq \!\{  x\!\in\!\mathbb{R}^{\xdim} \mid \forall \ctrl(\cdot),\;\;\exists s\!\in\![0, t], \;\textrm{s.t.}\; \traj(s)\!\in\!\targetset \}. \vspace{-0.25em}
\end{equation*}
If we consider an indefinite time horizon, we get the (minimal) time-invariant BRT, $\brt(\targetset):=\lim_{t \rightarrow \infty}\brt(t; \targetset)$.

Given the safety constraint set $\constraintset$, consider the BRT of the failure region $\constraintset^c$. Then, $\{\brt(\constraintset^c)\}^c$ is the maximal control invariant set contained in $\constraintset$. For any trajectory exiting this set, the trajectory will inevitably reach $\constraintset^c$ and therefore violate the constraint~\cite{Wabersich2023}. Thus, by solving the BRT problem, we determine the maximal safe set within the constraint set. We denote this set as $\safeset(\constraintset)$\footnote{More formally, $\safeset(\constraintset)\!\!:=\!\!\{\hspace{-1pt}\brt(\constraintset^c)\hspace{-1pt}\}^c$ is the \textit{viability kernel} of $\constraintset$~\cite{aubin2011viability}.}.

The BRAT is the set of initial states from which the system can reach a target set $\targetset$ while avoiding the failure region $\constraintset^c$ (or equivalently, while staying within the constraint $\constraintset$):
\vspace{-0.25em}
\begin{equation}
\begin{aligned}
    &\brat(t; \targetset, \constraintset^c) \! \coloneq \!  \left\{x \in \mathbb{R}^{\xdim} \mid \exists \ctrl(\cdot) \; \text{s.t.} \right. \\ 
    &\qquad \left. \exists s \in [0, t], \traj(s) \in \targetset \; \& \; \forall \tau \in [0, s], \traj(\tau) \in \constraintset  \right\}.
\end{aligned}     
\vspace{-0.25em}
\label{def:brat}
\end{equation}

The maximum safe set $\safeset(\constraintset)$ and the BRAT $\brat(t; \targetset, \constraintset^c)$ can be solved by defining the corresponding Hamilton-Jacobi (HJ) reachability optimal control problems. We first represent the constraint and the target sets as level sets of Lipschitz functions, $\constraintfcn(\cdot)$ and $\target(\cdot)$, such that \vspace{-0.25em}
\begin{equation}
    \constraintset = \{x \mid \constraintfcn(x) \ge 0 \}, \quad \targetset = \{x \mid \target(x) \ge 0 \}.
    \label{eq:unsafe-set}
    \vspace{-0.25em}
\end{equation}

Then, we can define the value functions whose zero-superlevel sets represent $\safeset(\constraintset)$ and $\brat(t; \targetset, \constraintset^c)$, respectively, as:

\hruleopen
\noindent Value function for $\safeset(\constraintset)$ and BRT:
\begin{align}
    V(x, t)  :&\!\!= \max_{\ctrl(\cdot)} \min_{s \in [0, t]}    \constraintfcn(\traj(s))  \label{eq:brt-value} \\ 
    & \Rightarrow \brt(t; \constraintset^c) =\{x \mid V(x, t) < 0 \}, \quad\quad\quad\quad \nonumber \\
    V(x)  :&\!\!= \lim_{t \rightarrow\infty} V(x, t)\quad \;\;\;\;~\; \label{eq:safe-value}\\ 
    & \Rightarrow \safeset(\constraintset) =\{x \mid V(x) \ge 0 \}. \quad\quad\quad\quad \nonumber
\end{align}
\noindent Value function for BRAT:
\begin{align}
    \label{eq:brat-value}
    W(x, t) :&\!\!= \max_{\ctrl(\cdot)} \max_{s \in [0, t]} \min \!\left\{ \target(\traj(s)), \min_{\tau \in [0, s] } \constraintfcn(\traj(\tau))\! \right\}\!\!\! \\
    & \Rightarrow \brat(t; \targetset, \constraintset^c) =\{x \mid W(x, t) \ge 0 \}. \nonumber
\end{align}
\hruleclose

\noindent The optimal control in \eqref{eq:brt-value} maximizes the closest distance to the failure boundary. In \eqref{eq:brat-value}, the optimal control minimizes the distance to the target set while avoiding the failure region. By applying the dynamic programming principle, the value functions become viscosity solutions~\cite{bardi1997optimal} to the HJ partial differential equations (PDEs), derived in~\cite{fisac2015reach}. The value functions can be computed numerically via the level set method~\cite{mitchell2002application} or approximated as neural value functions using physics-informed neural networks~\cite{bansal2021deepreach}.

\subsection{Control Barrier Function-based Safety Filtering}
\label{ssec:cbfs}

We address safety by using the computed BRT value function $V$ as control barrier functions (CBFs)~\cite{Ames2016}, to filter a potentially unsafe input $u$ with an appropriate safe input and keep the system within the constraint set $\constraintset$. 

To achieve this safety filtering mechanism, we consider the following min-norm safety filter~\cite{Wabersich2023, Ames2016}:
\hruleopen
\noindent CBF Min-norm Safety Filter:
\begin{subequations}
\label{eq:cbf-qp}
\begin{align}
u_{\mathrm{cbf}} =\; & \underset{u \in U}{\arg\min}  \quad \norm{u-u_{\mathrm{ref}}}^2 \\
\text{s.t.} \quad & \nabla V(x) \cdot f(x, u) +\alpha\left(V(x)\right) \geq 0,\label{eq:CBF-QP-constraint}
\end{align}
\end{subequations}
\hruleclose

\noindent where $\alpha$ is a class $\kappa$ function. For control-affine dynamics, this becomes the CBF quadratic program (CBF-QP). If $V$ is almost-everywhere differentiable, the barrier constraint in \eqref{eq:CBF-QP-constraint} is feasible almost-everywhere. Such usage of the reachability value function as the CBF is a prominent method of designing CBFs~\cite{choi2021robust, zhangdiscrete, oh2025safety}. For certain dynamics, the value function can be discontinuous without introducing a discount factor in time to the cost function in \eqref{eq:brt-value}~\cite{choi2023forward}. Various architectures of extending the safety filter \eqref{eq:cbf-qp} to our multi-agent problem will be discussed in Section~\ref{subsec:safety-fitering}.

\subsection{Time-To-Reach in Reachability Problems}
\label{ssec:TTR}
We consider the time-to-reach of the backward reach-avoid tube problem, given by
\begin{equation*}
T[\ctrl(\cdot)] \coloneq \begin{cases}
\infty \quad \text{if } \resizebox{0.66\hsize}{!}{$\displaystyle\exists t > 0,\;\text{s.t.}\;\traj(t) \in \constraintset^c \;\text{\&}\; \forall s\in [0, t], \traj(s) \notin \targetset$} \\
\infty \quad \text{else if}\; \forall t > 0, \traj(t) \notin \targetset, \\
\min \{ t \mid \traj(t) \in \targetset\} \quad \text{else.}
\end{cases}
\end{equation*}

In words, the $T[\cdot]$ of a given control signal is the time the trajectory first hits the target set without entering the failure region $\constraintset^c$. Next, we define the following optimal control problem, where the control tries to minimize $T$:
\vspace{-0.25em}
\begin{equation}
    \ttr(x):= \inf_{\ctrl(\cdot)} T[\ctrl(\cdot)],
\label{eq:min-ttr} \vspace{-0.5em}
\end{equation}
$\ttr(x)$ is called the \textit{minimum time-to-reach} (TTR) value function~\cite[Ch.2.3.1]{mitchell2002application}, in this case, for the backward reach-avoid tube problem.

\begin{proposition}
The BRAT of the time horizon $t$ and the TTR satisfy $\brat(t; \targetset, \constraintset^c) =\{x \mid \ttr(x) \le t \}$
\cite[Prop.1]{yang2013one}. \end{proposition}
The optimal control law of this problem is given as: \vspace{-0.25em}
\begin{equation}
    \pi^{\text{opt}}(x) = \mathrm{arg}\min_{u \in U} \nabla \ttr(x) \cdot f(x, u),
    \label{eq:ttr-opt-ctrl} \vspace{-0.25em}
\end{equation}
derived from the Hamiltonian associated with $\ttr$~\cite{yang2013one}. Note that the trajectory under this control law will be the time-optimal trajectory to $\targetset$ while avoiding $\constraintset^c$. Such a time-optimal trajectory always exists when $\ttr(x)$ is finite.

This minimum TTR value function-based control law will be used to design the control input that guides the vehicles to the air corridor in Section~\ref{subsec:ttr-separation}.

\begin{remark}
\label{remark:ttr}
Various numerical techniques are developed for the computation of the minimum TTR~\cite{mitchell2002application, doshi2022hamilton, yang2013one}.
Directly solving the stationary HJ-PDE associated with $\ttr$ is possible by carefully treating the discontinuity of $\ttr$~\cite[Ch.2.3.1]{mitchell2002application},~\cite{yang2013one}. In our work, we first compute the value function $W(x, t)$ and reconstruct $\ttr$ from $W$, as in~\cite{helperOC}. \end{remark}

\section{Problem Formulation}
\label{sec:prob_form}

\subsection{Problem Setting}

\begin{figure*}
\centering
\vspace{-0.5em}
\includegraphics[width=\textwidth]{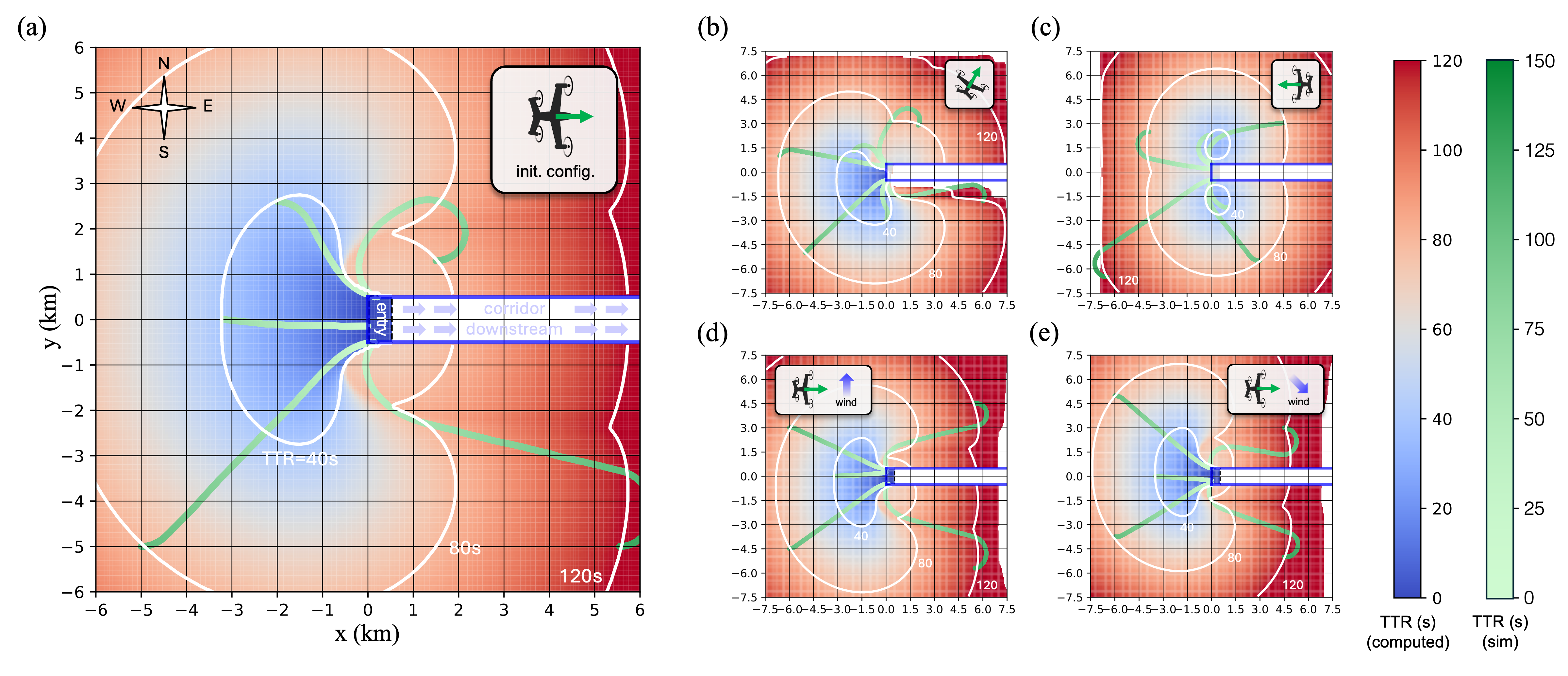}
\vspace{-2.5em}
 \caption{Computed minimum reach-avoid time-to-reach (TTR) in colormap (red-blue), for vehicles navigating to the air corridor (blue), and the simulated optimal trajectories (green). The corridor consists of the entry region $\targetset$ (black dashed), and the downstream region $\downstream$ (white background). \textbf{(a)} TTR colormap and optimal trajectories with initial heading $\theta=0^\circ$ (eastbound). \textbf{(b, c)} TTR colormap and optimal trajectories with initial heading $\theta=60^\circ, 180^\circ$, respectively. \textbf{(d, e)} TTR colormap and optimal trajectories with initial heading $\theta=0^\circ$, under constant wind blowing toward north and southeast, respectively.}
\label{fig:ttr}
\vspace{-1em}
\end{figure*}

We consider $N$ homogeneous aircraft approaching an air corridor where the $i$-th aircraft has dynamics
\[\dot{\traj}^i(t) = f(\traj^i(t), \ctrl^i(t))\]
where $\traj^i(t) \in \R^n$ is the vehicle's state and $\ctrl^i(t) \in U$ is their control. We define $[N] := \{1, \ldots, N\}$. 

The air corridor includes an entry region, which defines the target set $\targetset$ that each vehicle's state $\traj^i(t)$ aims to reach, and a downstream region $\downstream$ that vehicles must follow to remain within the corridor (Figure~\ref{fig:ttr} (a)). This downstream region $\downstream$ is treated as an avoid set
for any vehicle outside of the corridor, which prevents vehicles from entering the corridor from an unintended direction, e.g., 
perpendicular to the flow of traffic.
We assume that once a vehicle reaches the entry region, it proceeds along the corridor under a provided in-corridor guidance controller.

\subsection{Problem Statement}
\label{sec:statement}
For vehicles to safely enter the corridor, they must
(1) reach the entry region $\targetset$ while avoiding (2) (unintended) entry into the downstream region $\downstream$ and (3) conflicts with other vehicles.
The deconfliction objective is encoded as the constraint $\constraintset_{ij}:=\{(x^i, x^j) \; | \; \dist(x^i, x^j) \ge \delta \}$ for each vehicle pair $i$ and $j$, where $\dist(x^i, x^j)$ is the distance between the two vehicles.

We assume the initial configuration of vehicles is feasible, meaning there exists a set of compatible trajectories such that every vehicle reaches the corridor without violating safety:\vspace{-0.25em}
\begin{equation}
    \begin{aligned}
\label{assume:initialstate}
    \forall i & \in [N], \ \exists \ctrl^i(\cdot)\; \; \text{s.t.} \; \; \exists T^i > 0, \\ 
     & \resizebox{0.96\hsize}{!}{$\displaystyle\traj^i(T^i) \in \targetset, \text{and}\; \forall \tau\!\in\![0, T^i), \traj^i(\tau) \in \downstream^c, \traj^i(\tau) \in \constraintset_{ij} \;\forall j\!\neq\!i. $} \nonumber         
    \end{aligned}
    \vspace{-0.25em}
\end{equation}
Note that if we consider only a single vehicle (without considering deconfliction against other vehicles), each vehicle has to be initialized within $\brat(t; \targetset, \downstream)$. This requires $\ttr (x_0^i) < \infty$, the initial minimum TTR of vehicle $i$.

Our main objective is to find a valid joint control sequence $\bigtimes_{i=1}^N \ctrl^i(\cdot) \in \mathcal{U}^{[N]}$ that minimizes the corridor entry time of the last vehicle, $T^{\text{last}}$, while maintaining safety. This goal is captured in the following optimization problem:
\begin{subequations} \label{eq:problem}
\begin{align}
    & \quad \quad \quad \quad \quad \;\; \min_{\ctrl^i(\cdot) \in \mathcal{U} \ \forall i \in {[N]}} \quad T^{\text{last}} \label{eq:min_T_last}\\
    \text{s.t.} & \quad \dot{\traj}^i(t) = f(\traj^i(t), \ctrl^i(t)), \quad \\
    & \quad T^i \leq T^{\text{last}}, \quad T^i  := \min  \tau \;\; \text{s.t.} \ x^i(\tau) \in \targetset, \label{eq:problem:ttr}\\
    & \quad \traj^i(\tau) \in \downstream^c, \quad \forall \tau \in [0, T^i), \quad \forall i \in [N], \\
    & \quad \text{dist}(\traj^i(\tau), \traj^j(\tau))\!\geq\! \delta,\; \forall \tau\!\in\! [0, \min(T^i, T^j)) \nonumber\\
    & \qquad \qquad \qquad \qquad \qquad \forall i,j \in [N],  \ i \neq j \label{eq:pair-constraint}.
\end{align}
\end{subequations}

Directly solving~\eqref{eq:problem} suffers from several challenges.
First, it requires joint minimization over all $N$ vehicles, which scales poorly as the problem grows exponentially with vehicle count.
Further, a direct solution necessitates centralized control and
suffers from the complexity of multi-agency. Each pair of agents introduces a safety constraint~\eqref{eq:pair-constraint}, and these constraints may conflict with each other.
Therefore, careful treatment of these potentially conflicting constraints is required, which is nontrivial~\cite{AloorJ-RSS-25}. 
For these reasons, the direct solution is infeasible and does not generalize well.

Instead of solving~\eqref{eq:problem}, we employ decentralized optimal controllers with pairwise safety filters. To cope with this complexity, we relax the problem and
allow slack in the safety constraint \eqref{eq:pair-constraint}, aiming to minimize the safety violation rate as captured in the metric below, where $\mathds{1}$ is the indicator:
\[
\resizebox{0.98\hsize}{!}{$\displaystyle\nu_{\text{unsafe}} = \frac{1}{\sum_{k=1}^N T^k} \sum_{i=1}^N \int_0^{T^i} \max_{j \neq i} \mathds{1}\{(x^i(t), x^j(t)) \notin \constraintset_{ij}(t)\} dt.$}
\]
This measures the fraction of time that vehicles spend in a safety-violating configuration relative to the total travel time of all vehicles. %

We consider additional desiderata to represent notions of fairness. We desire that vehicles reach the corridor in the order corresponding to increasing initial TTR $\ttr^i:=\ttr(x_0^i)$,
exactly capturing first-come-first-served fairness.
The ``closest'' vehicle with the lowest initial TTR is granted the highest priority, and therefore it has the right-of-way unless safety is compromised.
Therefore, the initial TTR values naturally induce a desired priority order. Without loss of generality, assume that vehicle indices are assigned such that $\ttr^i \leq \ttr^j$ for all $i < j$, i.e., $i$ has higher priority.
By considering the order of the actual arrival sequence, we can calculate deviation from the desired priority order: \vspace{-0.25em}
\begin{equation*}
 \tau_K = \frac{2}{N(N-1)}\sum_{i,j \in [N], i < j} \!\!\!\! \text{sgn} \big(i-j \big) \;\text{sgn} \big( T^i - T^j \big), \vspace{-0.5em}
\end{equation*}
which is the closed form of Kendall's coefficient~\cite{Kendall}, a common statistic to capture variation from a desired ordering.

\section{TTR-Based Coordination Method}
\label{sec:TTRPriority}

\subsection{Method Overview}
\label{subsec:overview}

The key insight behind our method is that, under ideal coordination, vehicles would arrive at the corridor entry evenly spaced in time, and consequently, evenly spaced in distance along the approach path. This observation suggests that, rather than directly coordinating vehicle positions, one can coordinate their TTR values. By regulating TTR values to be evenly distributed rather than allowing vehicles to greedily minimize their individual TTRs, vehicles can naturally form a well-structured and safe arrival sequence. Building on this intuition, the proposed framework uses a layered structure: (1) a strategic guidance layer based on TTR separation, and (2) a tactical safety layer based on CBF safety filtering with priority awareness. 

The overall method consists of the precomputation of the necessary value functions, as well as the strategic guidance layer and the tactical safety layer for control:

\noindent \textit{1. Offline computation.} We first compute HJ reachability value functions offline. In particular, we compute:
\begin{itemize}
    \item The BRAT value function $W(x, t)$ and the associated TTR value function $\ttr(x)$ for $\brat(t; \targetset, \downstream)$. The TTR function is used to evaluate the priority order and to design the strategic guidance controller.
    \item A pairwise safety value function $V(x)$ defined over the relative dynamics of a pair of vehicles, which is used in the safety filter that maintains separation.
\end{itemize}

\noindent \textit{2. Strategic Layer---TTR separation guidance.} At the strategic level, each vehicle is assigned a reference TTR function based on the priority order
to evenly distribute vehicle trajectories across the TTR domain. The guidance controller then regulates each vehicle’s motion to increase or decrease its TTR to achieve the desired temporal spacing. 

\noindent \textit{3. Tactical Layer---Safety filtering.} At the tactical level, we employ CBF-based safety filtering to ensure separation when the strategic guidance commands would otherwise fail. This safety filter acts as a reactive mechanism that minimally modifies the guidance command only when necessary to maintain safety. We propose multiple architectures with varying levels of centralization and priority coordination.

\subsection{Offline Reachability Computation}
\label{ssec:offline}

\textit{Modeling.} In this study, we assume that other altitude ranges are reserved for different air operations or for last-resort collision avoidance~\cite{TCAS2011introduction, kochenderfer2012_ACAS}; therefore, we focus on horizontal deconfliction. For simplicity, we consider a planar (2D) top-down model and leave the extension to full 3D dynamics for future work.

We consider a reduced-order dynamic model of an autonomous air taxi given by
\begin{equation}
\begin{aligned}
 \dot{p}_x = v \cos \theta +  w & \cos \phi, \quad \dot{p}_y = v \sin \theta + w \sin \phi, \\
 & \dot{\theta} = \omega, \quad \dot{v} = \mathrm{a},
\end{aligned}
\label{eq:airtaxi}
\end{equation}
where the state is defined as $x = [p_x;\, p_y;\, \theta;\, v]$, representing position, heading, and speed. The control input is defined as $u = [\omega;\, \mathrm{a}]$, corresponding to the turn rate and longitudinal acceleration, respectively. The speed is constrained to the range $[v_{\min}, v_{\max}]$, and the control input bounds are defined as $\mathcal{U} = [-\omega_{\max}, \omega_{\max}] \times [a_{\min}, a_{\max}]$. Finally, we also incorporate a wind effect, characterized by the magnitude $w$ and the wind direction $\phi$. Note that the vehicle cannot stop in midair for maintaining wing-borne flight ($v_{\min} > 0$).

\begin{figure*}[h]
\centering
\vspace{-0.5em}
\includegraphics[width=0.8\textwidth]{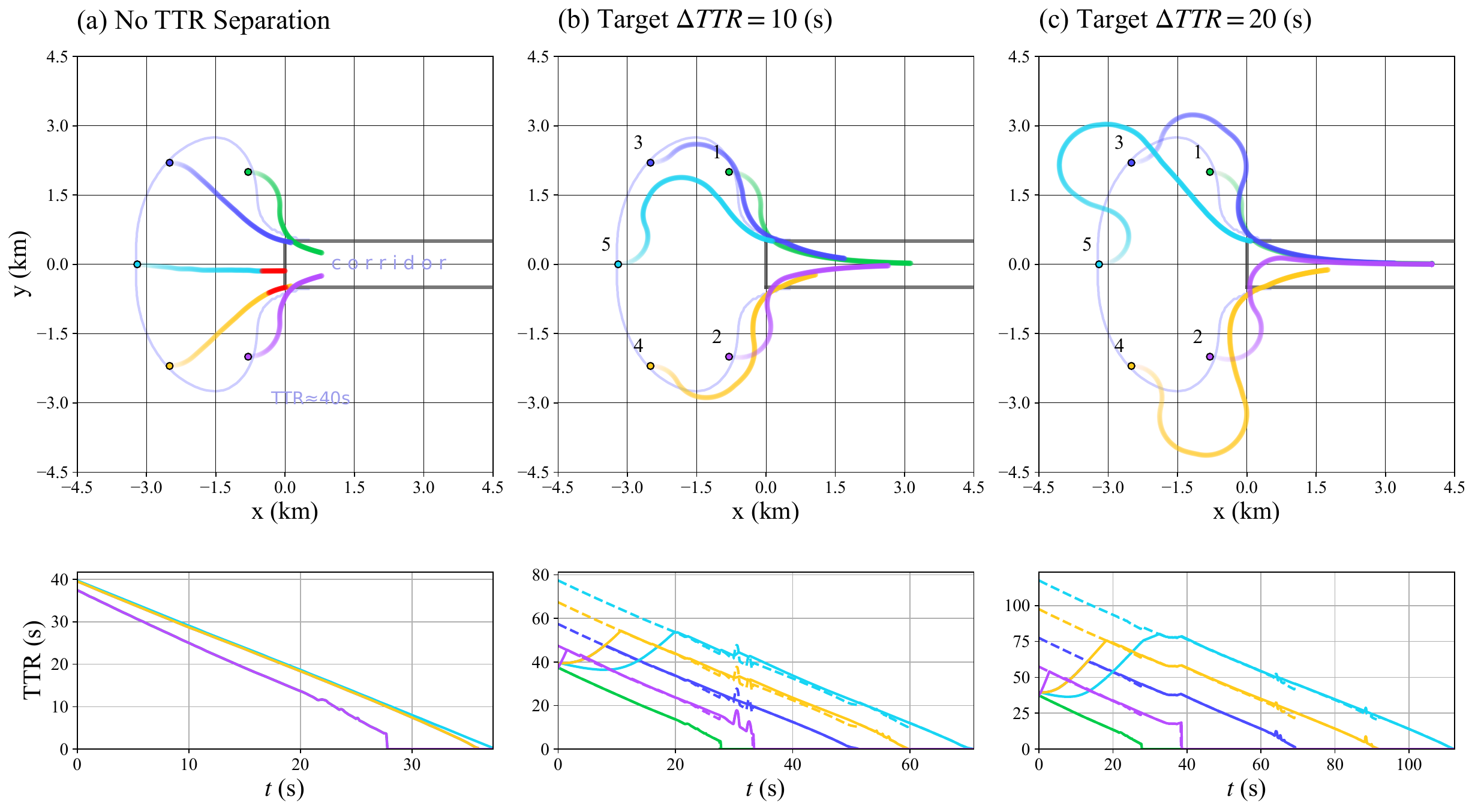}
\vspace{-1em}
 \caption{Five vehicles initialized at states with $TTR(x_i)\approx 40$s and with initial $\theta=0^\circ, v=v_{\scriptsize{\text{nominal}}}$. The first row shows the trajectories and the second row shows the TTR values over time (dashed line indicates target TTR values.) \textbf{(a)} Vehicles under TTR optimal control, leading to near-collision (red) near the corridor entry. \textbf{(b, c)} Vehicles under the TTR separation guidance with target separation time $\Delta T = 10, 20$s, respectively. The numbers indicate the assigned priority order. Note that in this scenario, the TTR separation guidance keeps the vehicles safe even without safety filtering.}
\label{fig:ttr_separation}
\vspace{-0.5em}
\end{figure*}

Without loss of generality, we assume that the corridor center is aligned with the $p_x$-axis and the origin is fixed at its entry. Let the width of the corridor be $d$. The corridor in the position space is defined as $\{(p_x, p_y) \mid p_x\! \ge \!0, |p_y|\!\le\!0.5d\}$. We design the entry region of the air corridor as the target set in the vehicle state space, given by
\begin{equation*}
    \resizebox{0.98\hsize}{!}{$\displaystyle \targetset \!=\!\{x \mid 0\!\le\!p_x\!\le\!0.5d,\; |p_y|\!\le 0.5d, \; |\theta|\!\le \Delta \theta,\; |v\!-\!v_{\text{nom}}|\!\le \Delta v \}. $}
\end{equation*}
The downstream region is defined as $\downstream\!=\!\{x \mid p_x\!\ge\!0.5d$, $|p_y|\!\le\!0.5d\}$. The parameters used in this work
are consistent with industry specifications~\cite{joby_aviation, archer_aircraft, wisk_aircraft, FAA_NMAC} and are
available in the accompanying source code and in \ifarxiv Table~\ref{tab:vehicle_dynamics} in the Appendix\else the Appendix of the full version\fi.

To define the relative dynamics for pairwise safety analysis, we use the relative state $x^{(ij)}\!=\![p_x^{(ij)};\,p_y^{(ij)};$ $ \, \theta^{(ij)};\, v^{i};\, v^{j}]$,
which includes the relative position and relative heading of vehicle $j$ with respect to vehicle $i$, expressed in a coordinate frame where the $p_x^{(ij)}$-axis is aligned with the heading of vehicle $i$. The relationship between the relative state and $(x^i, x^j)$, the relative dynamics, as well as visualizations of the computed $V(x^{(ij)})$, are detailed in~\cite{AloorJ-RSS-25}.

\vspace{0.25em}

\textit{Results.} The minimum TTR $\ttr(x)$ is computed per Remark~\ref{remark:ttr} and  presented in Fig.~\ref{fig:ttr}, along with the simulated time-optimal trajectories under the control law \eqref{eq:ttr-opt-ctrl}. In (a)-(c), we show slices of the value function at different heading angles. In (d) and (e), we introduce wind with speed $w\!=\!10$~m/s, and directions $\phi$ 
$=\!90^\circ, -45^\circ,$ respectively. Note that the presence of crosswind makes the TTR asymmetric with respect to the corridor centerline ($p_x$-axis), causing vehicles flying against the wind to arrive later than those flying with the wind. The reachability computation for the value functions $V(x^{(ij)}) $ and $ W(x, t)$ was completed within 30 minutes on an Nvidia RTX 5090 GPU.

\vspace{-0.25em}
\begin{remark}
\label{remark:numerical-error-TTR-gap}
Due to numerical discretization and approximation errors in the offline reachability computation, the computed TTR may differ from the actual TTR. In practice, we observe that this gap is less than approximately $10$ seconds. We find that our proposed method is robust to these slight numerical discrepancies.
\vspace{-3px}
\end{remark}

\vspace{-0.5em}
\subsection{TTR Separation Guidance}
\label{subsec:ttr-separation}

The objective of the TTR separation guidance is to coordinate multiple vehicles so that they arrive at the corridor entry temporally evenly separated.

\vspace{0.2em}

\textit{Priority assignment.} As specified in Section~\ref{sec:statement}, vehicles are assigned priority according to their initial TTR values. Once vehicles reach the corridor, they are removed from the priority ordering.

\vspace{0.2em}

\textit{Target TTR assignment.} 
The target TTR of the highest priority vehicle (i.e., index 1, the vehicle with the smallest TTR) is set equal to its current TTR: $\ttrtarget^{1}(0) := \ttr^1(0)$, and for all future times, $\ttrtarget^{1}(t) \coloneq \ttr(\traj^{1}(t))$. 
For the downstream vehicles with priority order $i > 1$, the target TTR values are assigned with respect to this TTR of the highest priority vehicle:
\vspace{-0.25em}
\begin{equation}
    \ttrtarget^{i}(t) = \ttrtarget^{1}(t) + (i-1)\Delta T,
\vspace{-0.25em}
\end{equation}
where $\Delta T$ is the desired temporal separation between consecutive vehicles.

\textit{Control law.} Each vehicle then regulates its motion to track its assigned target TTR value. The reference control for vehicle $i$ is defined as
\begin{align}
    \pi &^{\text{sep}, i} (\traj^i(t), t) =     \label{eq:ttr-separation-ctrl} \\
    & \resizebox{0.97\hsize}{!}{$\displaystyle\begin{cases}
    \underset{u \in U}{\max}\; \nabla \ttr(\traj^i (t)) \!\cdot\! f(\traj^i(t), u), & \;\; \text{if}\;\; \ttr(\traj^i (t)) \!\le \!\ttrtarget^{i}(t), \\
    \underset{u \in U}{\min}\;  \nabla \ttr(\traj^i (t)) \!\cdot\! f(\traj^i (t), u), & \;\; \text{else}.
    \end{cases}$} \nonumber
\end{align}
Vehicles with smaller-than-assigned TTR are guided to increase their TTR to maintain the target temporal separation, and vehicles with larger-than-assigned TTR are guided to reduce their TTR to follow the time-optimal trajectory to the corridor. This results in an even temporal spacing of vehicles.

\textit{Results.} Figure~\ref{fig:ttr_separation} illustrates the behavior of the proposed TTR separation guidance (without safety filtering), compared with greedy time-optimal guidance. In this scenario, five vehicles are initialized with TTR values near 40 seconds. %
Because TTR separation is enforced from the beginning of the maneuver, lower-priority vehicles intentionally deviate from their time-optimal trajectories, creating space for higher-priority vehicles to approach the corridor without interference. As a result, all vehicles enter the corridor in the correct order and without safety violations. This type of prescriptive coordination behavior is difficult to achieve using safety filtering alone, which primarily provides reactive conflict resolution. See Appendix for additional remarks.

\begin{table}[t]
\centering
\caption{Safety filter modes}
\vspace{-0.5em}
\label{tab:filter-modes}
\small
\renewcommand{\arraystretch}{1.15}
\resizebox{\columnwidth}{!}{%
\begin{tabular}{@{}lll@{}}
\toprule
Mode & Filter mechanism & Communication \\
\midrule
None & No safety filtering & -- \\
Dec & Pairwise CBF-QP; relaxed fallback & Neighbor states \\
Prio & Multi-stage priority; seq.\ propagation & Priority + Resolved $u$ \\
Cent & Joint CBF-QP over all vehicles & All states \& actions \\
\bottomrule
\end{tabular}%
}
\vspace{-1em}
\end{table}

\vspace*{-0.25em}
\subsection{Safety Filtering}
\label{subsec:safety-fitering}
\vspace{-0.25em}
Aligning safety priority with TTR priority is especially important under the guidance law in Section~\ref{subsec:ttr-separation} since unnecessary intervention on a higher-priority vehicle can create downstream delay and congestion for lower-priority vehicles.
We consider safety filters with varying levels of coordination, summarized in Table~\ref{tab:filter-modes}. The decentralized and centralized architectures have been considered in prior work~\cite{AloorJ-RSS-25, wang2017}. 
In this paper, we consider their combination with the greedy time-optimal reference control in \eqref{eq:ttr-opt-ctrl} as baseline methods for comparison in the results. 
\vspace{0.25em}

\noindent\textbf{Decentralized (Dec):}
Each vehicle independently solves a local pairwise CBF-QP against its most critical neighbor.
Only neighbor states are shared, and resolved control inputs are not propagated to other vehicles. For a vehicle pair $(i,j)$,
the \textit{most critical neighbor} for vehicle $i$ is
$j^*(i) = \arg\min_{j \in \mathcal{N}_i} V(x^{(ij)})$,
where $\mathcal{N}_i$ is the set of vehicles within the communication range of vehicle $i$. If the QP becomes infeasible, it finds a fallback solution with a relaxed CBF-QP, detailed in the Appendix.

\vspace{0.25em}
\noindent\textbf{Priority-Coordinated (Prio):}
Each vehicle solves a local pairwise CBF-QP incorporating the static priority order. Resolved controls are broadcast
and used as reference inputs in subsequent vehicles' CBF-QPs.  %
When determining the control input for each vehicle, we follow a multi-stage logic that is detailed below. 

Let $\mathcal{R}$ denote the set of vehicles whose controls have been resolved.
Each stage sweeps through vehicles in priority order. The currently selected ego vehicle $i$ attempts to resolve $u^i$ against its most critical neighbor, $j=j^*(i)$.

\noindent\textit{Stage~1} (priority-assigned): %
In Stage 1, the higher-priority vehicle's control is fixed to its reference input, thereby assigning the burden of conflict resolution to the lower-priority vehicle:
\hruleopen
Priority-assigned CBF-QP:
\begin{align}\label{eq:cbf-qp-priority}
  \min_{u^i \in U} \quad & \| u^i - u^i_{\mathrm{ref}} \|^2 \\
  \mathrm{s.t.} \quad
    & \nabla V(x^{(ij)}) \cdot f\!\bigl(x^{(ij)},\, u^i, u_{\mathrm{ref}}^j\bigr)
      + \alpha\, V(x^{(ij)}) \geq 0, \nonumber \\
    & u^i = u^{i}_{\mathrm{ref}} \; \text{if}\; i < j.\nonumber 
\end{align}
\hrule
\vspace{0.35em}

\noindent If vehicle $i$ has priority, this QP checks whether $u^i_{\mathrm{ref}}$ is feasible.
Otherwise, vehicle $i$ seeks an avoidance maneuver that accommodates $u^j_{\text{ref}}$.
A feasible solution resolves the control for vehicle $i$ (i.e., $\mathcal{R} \gets \mathcal{R} \cup \{i\}$) and is broadcast.

\noindent\textit{Stage~2} (symmetric):
For vehicles unresolved by Stage 1, we drop the fixed reference constraint and solve the symmetric QP~\eqref{eq:cbf-qp}.
If feasible with both vehicles applying avoidance, the ego vehicle commits and
$\mathcal{R} \gets \mathcal{R} \cup \{i\}$.

\noindent\textit{Stage~3} (relaxed symmetric):
For the remaining unresolved vehicles, we solve the relaxed version of the Stage 2 QP, available in the Appendix.
This stage is always feasible and ensures that all vehicles are resolved.

The last two stages are similar to the decentralized case, but the resolved control input is carried forward to inform the control selection for subsequent vehicles. This design does not require a centralized coordinator: vehicles publish their TTR values to neighboring agents, determine the priority order locally, and resolve the safe control input accordingly.

\vspace{0.5em}
\noindent\textbf{Centralized (Cent):} A single centralized QP in the Appendix jointly optimizes all vehicles' controls with pairwise CBF constraints coupling their decision variables. This safety filter requires the greatest degree of centralized infrastructure and vehicle coordination to simultaneously determine corrective actions for all vehicles.

\section{Simulation Results and Evaluation}
\label{ssec:results}
We evaluate different combinations of guidance and safety filtering across 100 randomly sampled $N\!=\!8$ vehicle scenarios.
Two guidance strategies are considered: (1) \textit{TTR~min}, where each vehicle greedily minimizes its TTR under the optimal control~\eqref{eq:ttr-opt-ctrl}, and (2) \textit{TTR~sep}, which applies the TTR separation guidance~\eqref{eq:ttr-separation-ctrl} with $\Delta T = 10$\,s. Each is combined with the safety filter modes in Table~\ref{tab:filter-modes}.

\begin{figure*}[t]
  \centering
  \includegraphics[width=0.95\textwidth]{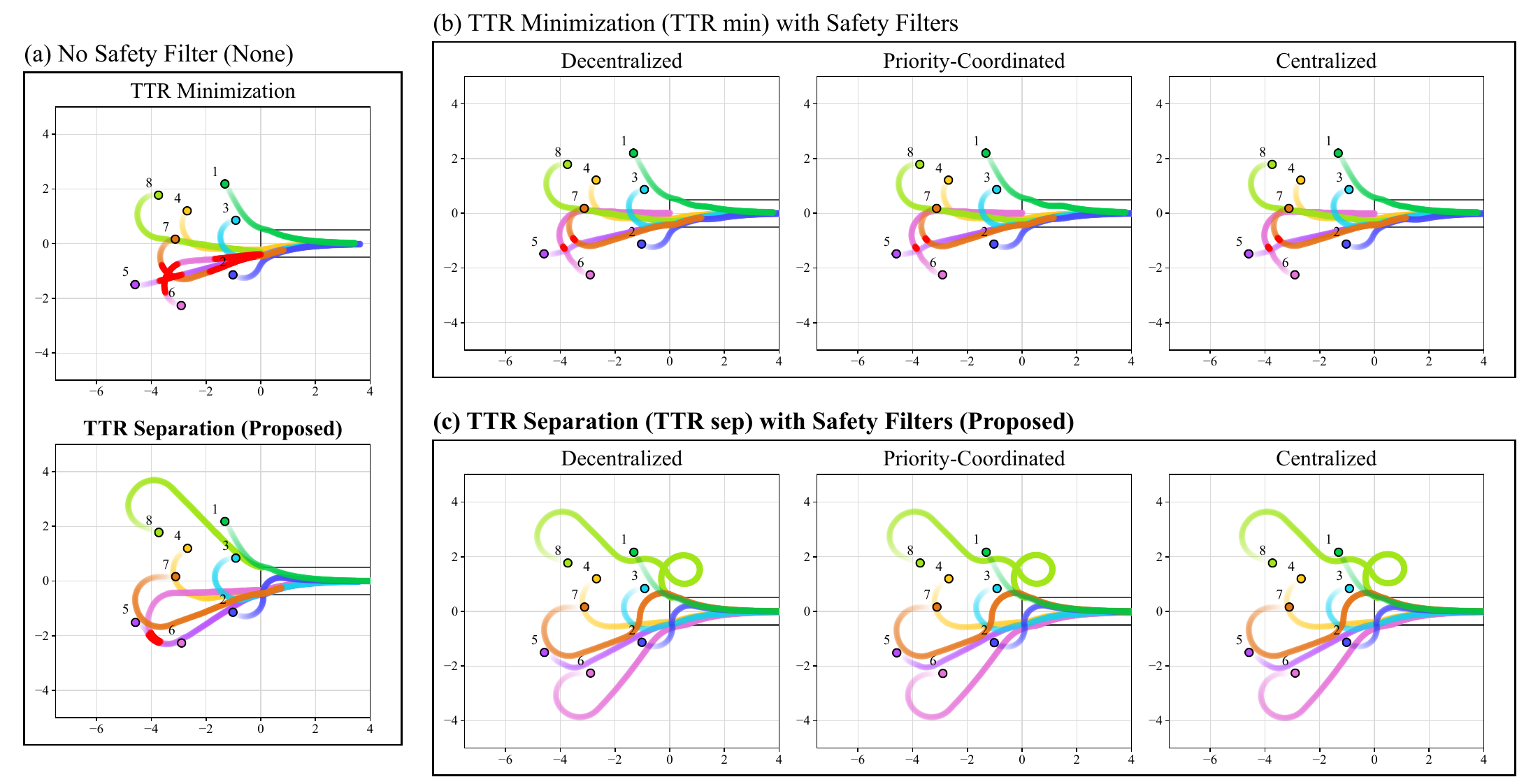}
  \vspace{-1em}
  \caption{Trajectories for each combination of reference guidance and safety filters on a representative scenario with eight vehicles. Red indicates safety violations. \textbf{(a)} Guidance without safety filtering: greedy TTR minimization leads to conflicts, while the proposed TTR separation reduces conflicts through temporal spacing.
\textbf{(b)} Greedy TTR minimization with safety filtering, which still incurs safety violations in this scenario.
(c) Proposed TTR separation combined with safety filtering achieves zero safety violations while maintaining the assigned arrival order. The animated trajectories of this scenario are available online at: \url{https://www.youtube.com/watch?v=c1ZX6twjodc}}
  \label{fig:trajectory-comparison}
\end{figure*}

\begin{table}[t]
\centering
\small
\caption{Evaluation metrics}
\vspace{-0.5em}
\label{tab:metric-defs}
\renewcommand{\arraystretch}{1.1}
\begin{tabular}{lp{0.7\columnwidth}}
\toprule
Symbol & Definition \\
\midrule
  $\nu_\text{unsafe}$ & System level safety (Sec~\ref{sec:statement}). \\
  $d_{\min}$ & Avg. minimum pairwise distance per run. \\
  $\tau_K$ & Kendall's tau (Sec~\ref{sec:statement}). \\
  $\text{std}(\sfrac{\ttrdelay}{\ttrtarget})$ & Std. of normalized $\ttrdelay^i$ normalized by $\ttrtarget^i$. \\
  $\ttrdelayptwo[\text{avg}]$ & Avg. non-negative $\ttrdelay^i$. \\
  $T^{\text{last}}$ & Actual arrival time of final vehicle. \\
  $P_\text{pass}$ & Success rate of $\nu_\text{unsafe}=0$ and $T^\text{last}<\infty$. \\
  $\Ppasstwo$ & $P_\text{pass}$ criteria and order maintained ($\tau_K = 1$). \\
  QP\textsubscript{time} & Safety filter wall-clock solve time per vehicle per timestep. \\
\bottomrule
\end{tabular}
\vspace{-1em}
\end{table}

\begin{table*}[t]
\centering
\caption{Evaluation of Proposed TTR Prioritization Control Scheme}
\label{tab:eval-results}
\adjustbox{max width=\textwidth}{%
\begin{tabular}{lllrrrrrrrrr}
\toprule
 &  &  & \multicolumn{2}{c}{Safety} & \multicolumn{2}{c}{Fairness} & \multicolumn{2}{c}{Efficiency} & \multicolumn{3}{c}{Overall} \\
\cmidrule(lr){4-5} \cmidrule(lr){6-7} \cmidrule(lr){8-9} \cmidrule(lr){10-12}
 & $u_\text{ref}$ & Filter & $\downarrow$~$\nu_{\text{unsafe}}$~($\%$) & $\uparrow$~$d_{\min}$~(km) & $\uparrow$~$\tau_K$ & $\downarrow$~$\text{std}(\sfrac{\ttrdelay}{\ttrtarget})$ & $\downarrow$~$\ttrdelayptwo[\text{avg}]$~(s) & $\downarrow$~$T^{\text{last}}$~(s) & $\uparrow$~$P_\text{pass}$ & $\uparrow$~$\Ppasstwo$ & $\downarrow$~QP\textsubscript{time}~(ms) \\
\midrule
\multirow{2}{*}{\rotatebox{0}{\scriptsize (a)}} & TTR min & None & $7.836 \pm 7.616$ & $0.313 \pm 0.216$ & $0.96 \pm 0.06$ & $0.000 \pm 0.000$ & $0.00 \pm 0.00$ & $77.77 \pm 6.88$ & $19.00$ & $8.00$ & -- \\
 & TTR sep & None & $2.081 \pm 2.667$ & $0.442 \pm 0.271$ & $0.99 \pm 0.02$ & $0.057 \pm 0.029$ & $3.47 \pm 2.45$ & $90.32 \pm 5.86$ & $43.00$ & $43.00$ & -- \\
\midrule
\multirow{3}{*}{\rotatebox{0}{\scriptsize (b)}} & TTR min & Dec & $0.194 \pm 0.744$ & $0.536 \pm 0.088$ & $0.91 \pm 0.12$ & $0.066 \pm 0.160$ & $1.23 \pm 2.42$ & $82.42 \pm 13.71$ & $90.00$ & $35.00$ & $4.1 \pm 0.1$ \\
 & TTR min & Prio & $0.256 \pm 0.876$ & $0.535 \pm 0.089$ & $0.91 \pm 0.12$ & $0.071 \pm 0.171$ & $1.48 \pm 3.31$ & $83.88 \pm 16.30$ & $87.00$ & $33.00$ & $5.2 \pm 0.3$ \\
 & TTR min & Cent & $0.742 \pm 1.594$ & $0.532 \pm 0.091$ & $0.91 \pm 0.13$ & $\mathbf{0.058 \pm 0.156}$ & $\mathbf{1.06 \pm 2.40}$ & $\mathbf{81.15 \pm 10.75}$ & $72.00$ & $32.00$ & $42.5 \pm 5.1$ \\
\midrule
\multirow{3}{*}{\rotatebox{0}{\scriptsize (c)}} & \textbf{TTR sep} & \textbf{Dec} & $0.118 \pm 0.634$ & $0.593 \pm 0.126$ & $\mathbf{0.99 \pm 0.03}$ & $0.101 \pm 0.111$ & $6.91 \pm 8.84$ & $97.67 \pm 17.89$ & $95.00$ & $\mathbf{82.00}$ & $\mathbf{4.0 \pm 0.1}$ \\
 & \textbf{TTR sep} & \textbf{Prio} & $0.058 \pm 0.327$ & $0.593 \pm 0.124$ & $\mathbf{0.99 \pm 0.03}$ & $0.105 \pm 0.118$ & $7.27 \pm 9.88$ & $99.69 \pm 23.68$ & $\mathbf{96.00}$ & $\mathbf{82.00}$ & $4.8 \pm 0.3$ \\
 & \textbf{TTR sep} & \textbf{Cent} & $\mathbf{0.044 \pm 0.184}$ & $\mathbf{0.593 \pm 0.128}$ & $\mathbf{0.99 \pm 0.03}$ & $0.094 \pm 0.100$ & $6.33 \pm 8.03$ & $96.16 \pm 15.25$ & $93.00$ & $80.00$ & $36.0 \pm 3.7$ \\
\bottomrule
\end{tabular}%
}
\par\vspace{2pt}\raggedright\footnotesize Simulation results from 100 randomly sampled scenarios. The best value for each metric is shown in \textbf{bold}. The first two rows do not treat safety and are excluded from this comparison. The solve time QP\textsubscript{time} is summed (per vehicle per timestep) over QPs in multi-stage when multiple stages were attempted. Experiments were run on an AMD 9800x3D CPU, CLARABEL~\cite{goulartClarabelInteriorpointSolver2026} was used to solve the QP.\vspace{-1.75em}
\end{table*}

\subsubsection{Evaluation Metrics}
The main metrics for efficiency, safety, and fairness are defined in Section~\ref{sec:statement}; they are
the travel time of the last vehicle arriving in the corridor $T^{\text{last}}$, the safety violation rate $\nu_{\text{unsafe}}$, and Kendall's coefficient ${\tau}_K$, respectively.  %
For each simulation run, any safety violation is considered a significant failure. Thus, we consider a run to be successful if all vehicles reach the goal with no safety violations. We record the rate of successful runs as $P_\text{pass}$. We also record the rate $\Ppasstwo$, which additionally requires that the arrival order perfectly conforms to the initial priority order. These rates encode both efficiency (goal-reaching) and safety objectives, with $\Ppasstwo$ further measuring fairness. They are interpreted as \emph{overall} metrics for success.
We summarize the full evaluation criteria in Table~\ref{tab:metric-defs}.

\subsubsection{Comparison Setup}
Table~\ref{tab:eval-results} presents the results for different combinations of guidance and safety filtering strategies. 
In (a), we evaluate guidance strategies without safety filtering, where \textit{TTR~min} serves as an uncoordinated baseline and \textit{TTR~sep} demonstrates the effect of temporal separation alone. 
In (b), we evaluate the three safety filtering architectures combined with \textit{TTR~min} guidance, which does not have strategic coordination. 
In (c), we combine the proposed \textit{TTR~sep} with the safety filtering architectures, representing the full proposed framework, and compare it against the baseline methods in (b). This structure allows us to isolate the effect of temporal separation, safety filtering, and their combination.

\subsubsection{Interpretation of Results and Discussion} 

In Table~\ref{tab:eval-results}, we observe that TTR separation combined with any of the safety filtering architectures in (c) outperforms the baseline combinations in terms of safety ($\nu_{\text{unsafe}}$ and $d_{\min}$), fairness ($\tau_K$), and the overall metrics ($P_\text{pass}$ and $\Ppasstwo$).
Although the baseline methods in (b) achieve better efficiency, this comes at the cost of worse safety.
We find that TTR separation strictly outperforms the baseline in all safety and overall pass rate metrics for a given filter.

Among the proposed safety filters in (c), the priority-coordinated filter achieves the highest safe mission success rate
and halves the safety violation rate of the decentralized filter.
It does so while resolving control in approximately 5 ms, making it suitable for real-time operation.
The centralized filter achieves the lowest average safety violation rate and best efficiency metrics.
This is achieved
by permitting more frequent, but less severe, safety violations and
by spreading the avoidance burden across all vehicles.
As a result, the centralized method trails
in the overall pass rates,
while also requiring significantly larger QP solve time.

We highlight a representative scenario in Fig.~\ref{fig:trajectory-comparison}, in which all baseline methods, including TTR separation guidance without safety filtering, fail to maintain safety. In contrast, all of the proposed methods in (c) successfully deconflict the vehicles and maintain priority order. %

Our results demonstrate that TTR separation is an extremely effective mechanism for safe coordination of multi-agent systems. These highlights the importance of strategic temporal coordination in preventing conflicts before reactive safety intervention becomes necessary, and moreover, the importance of aligning priority orders with individual agents' control and safety filtering mechanisms for the best results.

\section{Conclusion}

This paper shows the effectiveness of TTR-based guidance in achieving efficient vehicle trajectories while enabling preemptive deconfliction maneuvers in an autonomous traffic management system. The proposed TTR-based separation and guidance framework
{can be extended with more sophisticated tracking controllers to solve increasingly}
complex multi-agent coordination problems. Future work also includes improving 
the scalability of the numerical computation of TTR by leveraging, for example, 
learning-based value function representations, which have found recent success~\cite{bansal2021deepreach}.
\label{sec:conc}

\balance
\bibliographystyle{IEEEtran}
\bibliography{./IEEEfull,references}

\vspace{-0.5em}

\ifarxiv
\newpage
\section*{Appendix}

\subsection{Parameters Used in Simulation Study}

\begin{table}[h]
\centering
\footnotesize
\caption{Parameters of Air Taxi Dynamics \& Air Corridor}
\label{tab:vehicle_dynamics}
\begin{tabular}{p{4.5cm} l}
\toprule
\textbf{Parameter} & \textbf{Air taxi (Sim)} \\ 
\midrule
\textbf{Groundspeed} & \\ 
\quad Minimum speed \(v_{\min}\) & 60 knot (30 m/s) \\ 
\quad Maximum speed \(v_{\max}\) & 175 knot (90 m/s) \\ 
\quad Nominal cruise speed \(v_{\text{nom}}\) & 136 knot (70 m/s) \\ 
\midrule
\textbf{Acceleration} & \\
\quad \(a_{\min}\) & -3.3 ft/s$^2$ (-1.0 m/s$^2$) \\ 
\quad \(a_{\max}\) & 6.6 ft/s$^2$ (2.0 m/s$^2$) \\ 
\midrule
\textbf{Angular Rate ($\omega_{\max}$)} & 0.1 rad/s \\ 
\midrule
\textbf{Sampling Time} & 0.1 s \\ 
\midrule
\textbf{Corridor Geometry} & \\ 
\quad Width ($d$) & 0.62 miles (1 km) \\ 
\quad Entry heading threshold ($\Delta \theta$) & $\pm 20^\circ$ \\ 
\quad Entry speed threshold ($\Delta v$) & 38.9 knot (20 m/s) \\ 
\midrule
\textbf{Separation Distance ($\delta$)} & 1640 ft (500 m) \\ 
\bottomrule
\end{tabular}
\end{table}

For the simulation study, initial vehicle states were sampled uniformly from
$(p_x, p_y, \theta, v) \in [-5, -0.5] \times [-3, 3] \times [-\pi, \pi) \times \{v_{\text{nom}}\}$.
Rejection sampling is applied to scenarios to ensure that every pair of vehicles is
initially separated by at least $1$~km, and that no
three vehicles have initial TTR values all within $10$~s of one another.

\begin{table}[h]
\centering
\footnotesize
\caption{Parameters of Simulation \& Numerical Computation}
\label{tab:sim_params}
\begin{tabular}{p{4.5cm} l}
\toprule
\textbf{Parameter} & \textbf{Value} \\
\midrule
\textbf{Simulation Study} & \\
\quad Number of vehicles ($N$) & 8 \\
\quad Number of scenarios & 100 \\
\quad Target separation ($\Delta T$) & 10 s \\
\quad Simulation horizon & 200 s \\
\quad Communication range & 1.9 miles (3 km) \\
\midrule
\textbf{Safety Filter} & \\
\quad CBF rate ($\alpha$) & 1.0 \\
\quad Slack penalties ($\lambda$, $\varepsilon$) & 10, $10^{-3}$ \\
\quad CBF margin ($V \ge V_{\min}$) & 0.01 \\
\midrule
\textbf{TTR Value Function} & \\
\quad Domain $(p_x, p_y)$ & $[-7.5, 7.5]^2$ km \\
\quad Domain $(\theta, v)$ & $[-\pi, \pi] \times [10, 110]$ m/s \\
\quad Grid size & $271 \times 271 \times 60 \times 21$ \\
\quad Horizon & 360 s \\
\midrule
\textbf{Pairwise Safety Value Function} & \\
\quad Domain $(p_x^{(ij)}, p_y^{(ij)})$ & $[-3, 3]^2$ km \\
\quad Domain $(\theta^{(ij)}, v^i, v^j)$ & $[-\pi, \pi] \times [30, 90]^2$ m/s \\
\quad Grid size & $121 \times 121 \times 32 \times 10 \times 10$ \\
\quad Horizon & 100 s \\
\bottomrule
\end{tabular}
\vspace{-1.5em}
\end{table}

\subsection{Additional remarks on TTR Separation Guidance}

\begin{remark}[Selection of $\Delta T$]
The minimum feasible value of $\Delta T$ for safety can be approximated by the required separation distance divided by the nominal vehicle speed. Under the parameters in Table~\ref{tab:vehicle_dynamics}, this corresponds to approximately $7.1$s. In practice, $\Delta T$ can be adjusted based on operational requirements, such as regulatory separation rules or vertiport throughput limits, and it can be used to handle numerical errors in the TTR computation. %
\end{remark}

\begin{remark}[Protocols for Implementation] %
For decentralized implementation, vehicles share their TTR values with neighboring agents and determine priority ordering within their communication range. For centralized implementation, a central traffic controller issues the target TTR values to each vehicle, and each vehicle executes the control law \eqref{eq:ttr-separation-ctrl}.
\end{remark}

\subsection{Relaxed QPs in Decentralized / Centralized Safety Filters}

\hruleopen
\noindent Decentralized Relaxed CBF-QP for vehicle $i$:
\begin{equation}\label{eq:pairwise-relaxed}
\begin{aligned}
  \min_{u^{(ij)},\,\xi} \quad & \| u^i - u_{\mathrm{ref}}^i \|^2
    + \lambda\, \xi + \tfrac{\varepsilon}{2}\, \xi^2 \\
  \mathrm{s.t.} \quad
    & \resizebox{0.87\hsize}{!}{$\displaystyle\nabla V(x^{(ij)})^\top f\!\bigl(x^{(ij)},\, u^i, u^j\bigr)
      + \alpha\, V(x^{(ij)}) + \xi \geq 0,\quad \xi \geq 0$}, \nonumber
\end{aligned}
\end{equation}
\hruleclose
\noindent where $\alpha > 0$ is a linear class-$\kappa$ function parameter consistent with~\eqref{eq:CBF-QP-constraint}. This relaxed QP is always feasible since $\xi$ is unbounded above.

\vspace{1em}

\hruleopen
\noindent Centralized CBF-QP:
\begin{equation}\label{eq:centralized}
\begin{aligned}
  \min_{u^{[N]},\,\xi^{[N]}} \quad & \sum_{i=1}^{N} \| u^i - u^i_{\mathrm{ref}} \|^2 + \lambda \!\sum_{(i,j) \in \mathcal{P}} \xi_{ij} + \tfrac{\varepsilon}{2} \!\sum_{(i,j) \in \mathcal{P}} \xi_{ij}^2 \\
  \mathrm{s.t.} \quad & \begin{aligned}[t]
    & \forall\, i, \quad u^i \in U &&  \\
    & \forall\, (i,j) \in \mathcal{P}, \\
    & \quad \xi_{ij} \geq 0, &&  \\
    & \quad \resizebox{0.75\hsize}{!}{$\displaystyle\nabla V(x^{(ij)})^\top f\!\bigl(x^{(ij)},\, u^{(ij)}\bigr) + \alpha\, V(x^{(ij)}) + \xi_{ij} \geq 0,$} \nonumber
  \end{aligned}
\end{aligned}
\end{equation}
\hruleclose
\noindent where we denote the set of active vehicle pairs as $\mathcal{P} := \{(i,j) \mid i < j, \; j \in \mathcal{N}_i\}$. 
$u^{[N]} \in \mathbb{R}^{N\udim}$ is the stacked control vector and $\xi^{[N]} \in \mathbb{R}^{|\mathcal{P}|}$ contains independent per-pair slack variables. The QP is relaxed for feasibility.

\fi

\end{document}